# 3D curvature analysis of seismic waveforms and its interpretational implications

Haibin Di[1,2], Motaz Alfarraj[1], and Ghassan AlRegib[1]

[1] Georgia Institute of Technology, School of Electrical and Computer Engineering, Center for Energy and Geo Processing (CeGP) at Georgia Tech and KFUPM, Atlanta, Georgia, USA.

[2] Schlumberger, Houston, Texas, USA.

E-mail: hdi7@gatech.edu; motaz@gatech.edu; alregib@gatech.edu

# ABSTRACT

The idea of curvature analysis has been widely used in subsurface structure interpretation from three-dimensional (3D) seismic data (e.g., fault/fracture detection and geomorphology delineation) by measuring the lateral changes in the geometry of seismic events. However, such geometric curvature utilizes only the kinematic information (two-way traveltime) of the available seismic signals. While analyzing the dynamic information (waveform), the traditional approaches (e.g., complex trace analysis) are often trace-wise and thereby fail to take into account the seismic reflector continuity and deviate from the true direction of geologic deposition, especially for steeply dipping formations. This study proposes extending the 3D curvature analysis to the waveforms in a seismic profile, here denoted as the waveform curvature, and investigates the associated implications for assisting seismic interpretation. Applications to the F3 seismic dataset over the Netherlands North Sea demonstrate the added values of the proposed waveform curvature analysis in four aspects. First, the capability of the curvature operator in differentiating convex and concave bending allows automatic decomposition of a seismic image by the reflector types (peaks, troughs and zero-crossings), which can greatly facilitate computer-aided horizon interpretation and modelling from 3D seismic data. Second, the signed minimum curvature offers a new analytical approach for estimating the fundamental and important reflector dip attribute by searching the orientation associated with least waveform variation. Third, the signed maximum curvature makes it possible to analyze the seismic signals along the normal direction of the reflection events. Finally, the curvature analysis promotes the frequency bands of the seismic signals and thereby enhances the apparent resolution on identifying and interpreting subtle seismic features.

# INTRODUCTION

The curvature analysis, as a fundamental tool of signal processing, has been widely used in various industries and disciplines, e.g., medical brain scanners (Joshi *et al*. 1995), optometry (Daniel and Barsky 1997), and terrain analysis (Wood 1996). Since its first introduction to subsurface structure mapping and hydrocarbon reservoir characterization (Lisle 1994), lots of efforts have been devoted to developing and implementing the curvature attribute as well as its variations to structure interpretation from three-dimensional (3D) seismic surveys (e.g., Roberts 2001; Bergler *et al.* 2002; Hart, Pearson and Rawling 2002; Bergbauer, Mukerji and Hennings 2003; Hart and Sagan 2005; Al-Dossary and Marfurt 2006; Blumentritt, Marfurt and Sullivan 2006; Hart 2006; Sullivan *et al.* 2006; Chopra and Marfurt 2010, 2013; Di and Gao 2014a, 2014b; Silva *et al.* 2016; Alfarraj, Di and AlRegib 2017).

Traditionally, the seismic curvature measures the second-order lateral changes in the geometry of seismic reflectors, which successfully depicts the surface morphology of rock layers and more importantly highlights the potential faults and fractures caused by anticlinal bending (Roberts 2001; Al-Dossary and Marfurt 2006; Gao 2013). Here we denote it as geometric curvature analysis for distinguishing the waveform curvature proposed in this paper. However, the reflector geometry represents only the kinematics (e.g., two-way traveltime) of the seismic waveforms but ignores the dynamic information (e.g., amplitude). In some cases, the latter is of more importance for understanding the seismic signals, particularly texture analysis, facies mapping, and amplitude interpretation (e.g., Widess 1973; Sen and Stoffa 1991; Crase *et al.* 1992; Partyka, Gridley and Lopez 1999; Pratt 1999; Neff, Runnestrand and Butler 2001; Coleou *et al.* 2003; Sirgue and Pratt 2004; Zeng 2004; Gao 2011, 2012; Di and Gao 2017). For robust waveform analysis, two conventional techniques are the complex seismic trace analysis (Taner, Koehler and Sheriff 1979;

Robertson and Nogami 1984; Barnes 1991) and the spectral decomposition (Claerbout 1976; Chakraborty and Okaya 1995; Peyton, Bottjer and Partyka 1998; Partyka *et al.* 1999; Marfurt and Kirlin 2001). Specifically, the former treats the reflection amplitude as the real part of a complex trace and performs the Hilbert transform to predict the quadrature amplitude. The complex-trace analysis has served as the foundation for extracting a set of popular seismic attributes, such as instantaneous phase/frequency and reflector dip/azimuth since its introduction in the 1970s (Taner *et al.* 1979; Barnes 1996). The latter converts the seismic signals into the frequency domain, so that many features that are difficult to visualize in the time domain becomes visible. Spectral decomposition is now widely used as a routine process when interpreting seismic data, such as thickness prediction and thin-bed analysis (Partyka *et al.* 1999; Marfurt and Kirlin 2001; Liu and Marfurt 2007). However, these existing methods often analyze seismic signals simply along the time/depth direction, instead of perpendicular to the reflectors and the associated rock layers. Therefore, such simplification may deviate the analysis from the true direction of geologic deposition and runs the risk of introducing artifacts, especially in the steeply-dipping formations.

Considering the capability of curvature analysis in finding the normal vector to any given curve/surface as well as differentiating the concave/convex components, this study proposes extending the traditional curvature operator to the seismic waveforms, here denoted as the waveform curvature, and more importantly investigating the associated implications for assisting seismic interpretation and modelling. The paper is structured as follows: First, we illustrate the differences between the traditional geometric curvature and the proposed waveform curvature. Second, we demonstrate the interpretational implications through applications to the F3 dataset over the Netherlands North Sea. Then, we provide an example of how the proposed waveform curvature analysis assists horizon modelling. Finally, we discuss the limitations to avoid potential misinterpretation of the generated waveform-curvature maps. Based on the results, we draw conclusions at the end of the paper.

## WAVEFORM CURVATURE

### Curvature in general

First, for the convenience of illustrating the differences between the traditional geometric curvature and the proposed waveform curvature, we rewrite the concept and equation to compute curvature in a more general way, instead of the conventional spatial coordinate system ($x$-$y$-$z$), to avoid potential misunderstanding and confusion in this paper.

In mathematics, the curvature describes how much a curve deviates from being a straight line, or a surface from being a flat plane, with its magnitude quantifying the degree of bending and its sign differentiating convex bending and concave bending. Take a two-dimensional (2D) curve for example (Fig. 1). One convenient way for understanding the curvature at point $P$ is to find the unique circle (or the osculating circle) that most-closely approximates the curve near $P$. The associated curvature is defined as the reciprocal of the circle radius at $P$. Therefore, the more the curve bends, the smaller the osculating circle and, and correspondingly, the larger the curvature.

The computation of curvature differs from 2D curves to 3D surfaces. Specifically, the solution is unique at a given point on a curve in 2D space. Let the curve be represented as function $f = f(x)$, the associated 2D curvature $k_{2D}(x, f)$ is evaluated as

$$k_{2D}(x, f) = \frac{f''}{\left(1+f'^2\right)^{3/2}}, \qquad (1)$$

in which $x$ denotes the coordinate along the horizontal axis of the 2D space. $f' = \frac{df}{dx}$ and $f'' = \frac{d^2f}{dx^2}$ denote the first and second derivative of $f$ with respect to axis $x$, respectively (Roberts, 2001).

When extended to a 3D surface, however, the solution becomes non-unique. There exist an infinite number of curvatures at any point on it, depending on how to perpendicularly cut the surface with a plane. Let the surface be represented as function $f = f(x, y)$, the associated 3D curvature $k_{3D}(x, y, f)$ turns to be dependent on the orientation $\varphi$ of the cutting plane, and the azimuth-dependency is described as

$$k_{3D}^{\varphi}(x,y,f) = \frac{1}{[1+A_1^2+A_2^2]^{1/2}} \cdot \frac{(B_1 + B_2 \tan^2\varphi + 2B_3 \tan\varphi)}{[(1+A_1^2)+(1+A_2^2)\tan^2\varphi + 2A_1 A_2 \tan\varphi]}. \qquad (2)$$

where $x$ and $y$ denote the coordinates along the two horizontal axes of the 3D space. $A_1 = \frac{df}{dx}$ and $A_2 = \frac{df}{dy}$ denote the first derivative of $f$ with aspect to axes $x$ and $y$, respectively. $B_1 = \frac{d^2 f}{dx^2}$, $B_2 = \frac{d^2 f}{dy^2}$ and $B_3 = \frac{d^2 f}{dxdy}$ denote the second derivatives (Di and Gao 2016a).

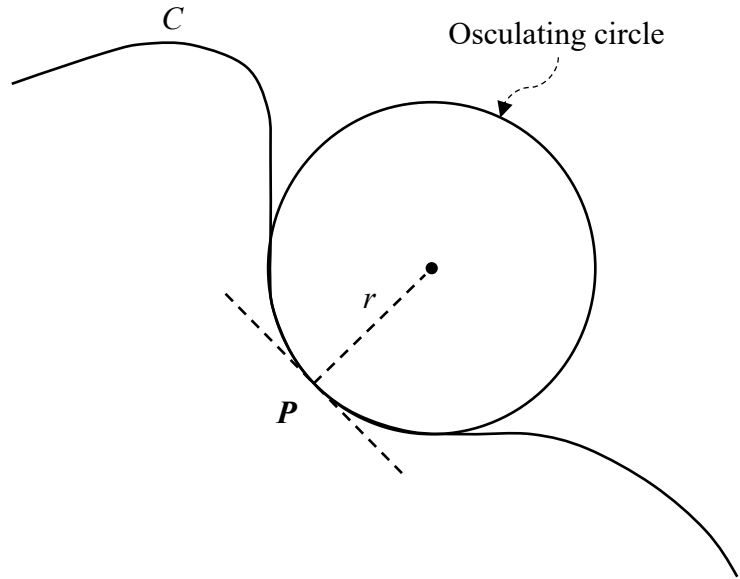

**Fig. 1. The sketch map for understanding the curvature of a curve. At point P on a curve C, there exists one unique circle (or the osculating circle) that most-closely approximates the curve near P. The associated curvature is defined as the reciprocal of the radius r of the osculating circle. (Modified from Cepheus 2006)**

The general $x$-$y$-$f$ definition varies by the different applications of curvature analysis, such as string vibration and fluid flow. Since its first introduction to seismic interpretation by Lisle (1994), the curvature analysis is typically applied for subsurface structure mapping and geomorphology delineation by quantifying the bending of seismic reflectors, which uses the $x$-axis representing the inline, the $y$-axis representing the crossline, and the $f$-axis representing the elevation or depth ($z$), also commonly known as the geospatial coordinate system ($x$-$y$-$z$) in most literatures.

**Geometric curvature vs. waveform curvature**

In general, 3D seismic surveying records the response of the rock layers when a wave propagates through them, and thereby two sets of information are stored at every sample, including its spatial location (inline, crossline, and depth/time) and the amplitude of reflection waveform. For the convenience of description in this paper, in addition to the $x$-$y$-$z$ geospatial system, we use $w$ to represent the waveform amplitude. Therefore, there exist three planes available for curvature analysis in math, one horizontal plane ($x$-$y$) and two vertical planes ($x$-$z$ and $y$-$z$). Among them, the curvature analysis in the horizontal plane has been well investigated for the purpose of structure interpretation and fault delineation and is popular in real applications (e.g., Roberts 2001). Such curvature could be further divided into two subcategories, with one being the structural curvature $k_{3D}(x, y, z)$ that evaluates the lateral variation of the reflector location $z$, and the other being the amplitude curvature $k_{3D}(x, y, w)$ that evaluates the lateral variation of the waveform amplitude $w$ (Chopra and Marfurt 2013). In this paper, we denote both curvature analyses in the horizontal $x$-$y$ plane as geometric curvature.

In math, the 2D/3D curvature analysis is also applicable to the waveforms in the vertical $x$-$z$ and $y$-$z$ planes, which is denoted as waveform curvature in this paper (Di, Alfarraj and AlRegib 2017). Take the $x$-$z$ plane for example. For illustrating the values of the proposed waveform curvature in seismic interpretation, we first calculate the 2D curvature $k_{2D}(z, w)$ (eq. 1) for a single trace randomly retrieved from the F3 seismic volume over the Netherlands North Sea. Fig. 2 compares the original seismic signal (in blue) and the calculated waveform curvature (in red) after normalization. Two observations are obtained. First, the waveform curvature matches the original signal very well, with all peaks and troughs clearly captured. Second, as the second-order derivative, the curvature operator acts like a high-pass filter and thereby enhances the visibility of the subtle features associated with low reflection amplitude (denoted as circles). Such enhancement in the apparent resolution is particularly helpful for seismic interpretation at small scale, such as sequence analysis, thickness prediction, and interbed delineation.

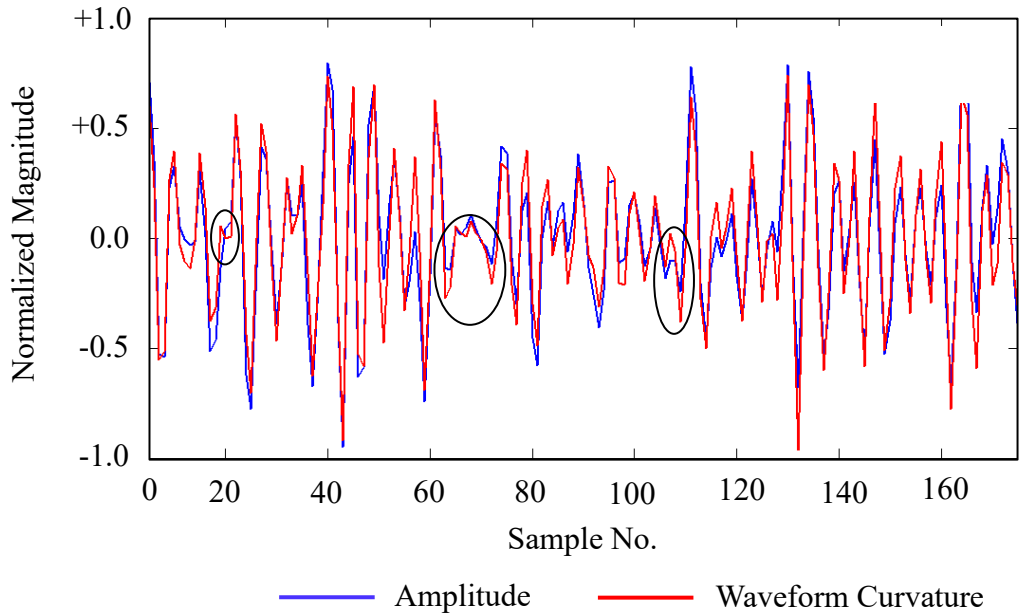

**Fig. 2. The 2D curvature analysis for a seismic trace randomly retrieved from the F3 dataset over the Netherlands North Sea. Both curves are normalized for fair comparison. Note the good match between the original waveform and the curvature, as well as the enhanced visibility of the subtle features associated with low reflection amplitude (denoted by circles). A time window of 3 samples is used for computation.**

More importantly, when the 3D curvature analysis (eq. 2) is applied to a seismic section, the associated waveform curvature $k_{3D}(x, z, w)$ quantifies how the reflection varies in the section and offers more insights for interpreting the seismic signals (Fig. 3). For example, at any given sample (denoted as orange dot), the most significant variation is observed when tracking it perpendicularly to the reflector (denoted by the blue arrow), whereas the waveform of a reflector is expected to vary least when tracking it along the reflector (denoted by the red arrow). Therefore, the former allows describing the seismic signals along the normal vector to the reflectors, which follows the true direction of geologic deposition and thereby most-physically represents the geology. The latter offers a new approach for predicting the direction of least waveform variation, or maximum reflector continuity, in the vertical section, which is equivalent to the dip attribute popularly used in the structure interpretation from seismic data (Di and AlRegib 2017). The comparisons between the traditional geometric curvature and the proposed waveform curvature are summarized in Tab. 1, and we discuss about the associated interpretational implications exclusively in the following section.

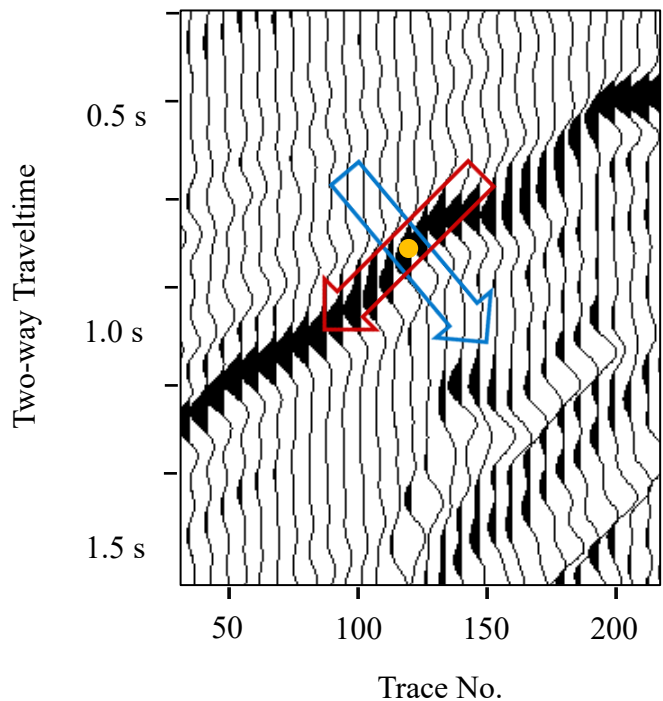

**Fig. 3. The diagram for illustrating the implications of performing 3D curvature analysis on a seismic section (Di et al. 2017). At a sample (denoted as orange dot), the signed maximum curvature describes the seismic signals perpendicularly to the reflectors (denoted by the blue arrow), which follows the true direction of geologic deposition and thereby is most physically linked to the geology. The signed minimum curvature predicts the direction of maximum reflector continuity in the vertical section (denoted by the red arrow), which provides a new approach for computing the dip attribute popularly used in structure interpretation.**

Table. 1. The comparisons between the traditional geometric curvature and the proposed waveform curvature analysis of 3D seismic data.

|  | **Geometric curvature** | **Waveform curvature** |
|---|---|---|
| Object | Curvature analysis for seismic signals | |
| Information for analysis | a. Structural depth/time $z$, or<br>b. Reflection amplitude $w$ | Reflection amplitude $w$ |
| Computation direction | Horizontal<br>(the $x$-$y$ plan) | Vertical<br>(the $x$-$z$ or $y$-$z$ plane) |
| Interpretational applications | Structure analysis:<br>a. Reflector morphology delineation<br>b. Fault detection<br>c. Fault characterization | Waveform analysis:<br>a. Reflector decomposition<br>b. Dip estimation<br>c. Apparent resolution enhancement |

## INTERPRETATIONAL IMPLICATIONS

For demonstrating the implications of the proposed waveform curvature analysis on seismic interpretation, we use a subset of the post-stack time-migrated F3 seismic dataset over the Netherlands North Sea, which consists of 451 inlines and 401 crosslines with 25 m as the bin size and 76 samples per trace with 2 *ms* as the sampling rate. Fig. 4 displays the post-stack amplitude in the vertical section of crossline #625, in which we notice a set of subparallel reflectors with varying reflection intensity. As listed in Tab. 1, the waveform curvature analysis could assist seismic interpretation in three aspects: reflector decomposition, dip estimation, and apparent resolution enhancement. To be clear, the proposed waveform curvature is calculated from the entire seismic volume with a window size of 4 *ms* vertically and 3 traces laterally. After computation, the results are then clipped to the same vertical section for visualization and analysis.

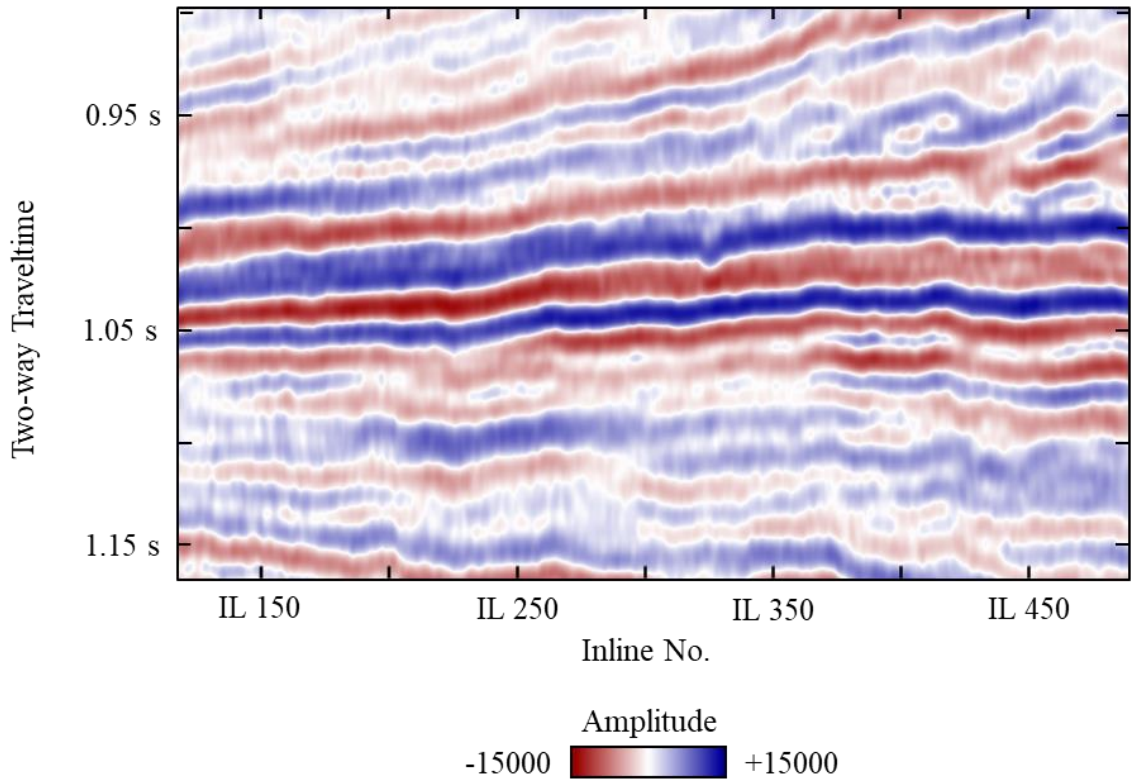

**Fig. 4. The seismic amplitude in the vertical section of crossline #625 from the F3 seismic dataset over the Netherlands North Sea for demonstrating the implications of the proposed waveform curvature analysis on assisting seismic interpretation.**

**Reflector decomposition**

In seismic interpretation, there are three major reflector types of great interpretational interest and important geologic implication: the peaks, troughs and zero-crossings. Practically, reflector interpretation in a given seismic dataset often focuses on a certain type depending on the wavelet phase. Automated separation of the target reflector type from the others could be helpful for computer-aided horizon interpretation without the interference from the surroundings reflection events of no interest. As demonstrated in Fig. 2, by simply treating a waveform as a curve of convex and concave components, performing the curvature analysis makes it possible for differentiating the waveform peaks and troughs using positive curvatures and negative curvatures, respectively. Correspondingly, the waveform is decomposed into two parts, with one for the peak reflectors and the other for the trough reflectors. Moreover, by integrating such decomposition method with the complex trace analysis, the zero-crossings could also be decomposed from the quadrature amplitude in a similar way, including the peak-over-troughs and the trough-over-peaks, often denoted as s-crossings and z-crossings. Fig. 5 displays the four types of reflectors decomposed from the vertical section of crossline #625, each of which is shown as a separate image. In addition, the curvature magnitude preserves the information about the reflection amplitude. The higher the waveform amplitude, typically the larger the curvature magnitude. Therefore, besides the successful decomposition of the reflection events, the proposed waveform curvature is also indicative of the contrast in the reflection intensity associated with rock layers of varying properties, both of which are essential for reliable horizon interpretation and modelling. More details are provided in the next section.

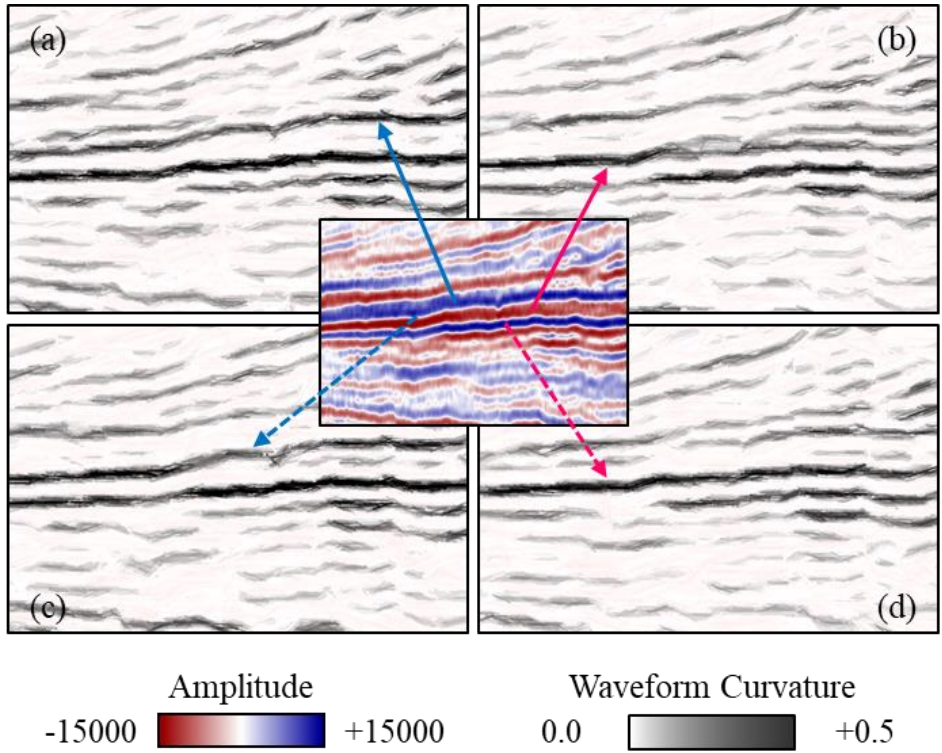

**Fig. 5. The decomposition of the four reflector types in the vertical section of crossline #625 by the proposed waveform curvature analysis, including the peaks (a), troughs (b), s-crossings (c), and z-crossings (d). The absolute waveform curvature is used for visualization here.**

**Dip estimation**

As illustrated in Fig. 3, the signed minimum curvature is associated with the least waveform variation, and thereby its orientation represents the direction of maximum reflector continuity, implying its potential for estimating the dip angle of seismic reflectors. Compared to the conventional methods of dip estimation such as complex-seismic-trace analysis (Barnes 1996), plane-wave destruction (Fomel 2002), and discrete scanning (Marfurt 2006), it is superior in two ways. First, it takes into account the full information about the local waveform, including amplitude, phase and frequency, which is thereby more robust to seismic noises, compared to the instantaneous phase-based method. Second, the curvature analysis has an analytical solution, which is thereby more computational efficient, compared to the computationally expensive discrete-scanning method. Fig. 6 displays the reflector dip estimates in the vertical section of crossline #625 by the proposed waveform curvature analysis, with the positive values for downward dipping (in blue) and the negative values for upward dipping (in red). As reflected in the dip map, all reflectors in the section dip slightly, and apparently, the upward dipping is observed in the upper area, whereas the positive dip indicates downward dipping in the lower area. Among all curvature computation algorithm (Al-Dossary and Marfurt 2006; Di and Gao 2016a, 2016b), the signed minimum curvature as well as its orientation is computed by using the surface-rotation based scheme (Di and Gao 2016a).

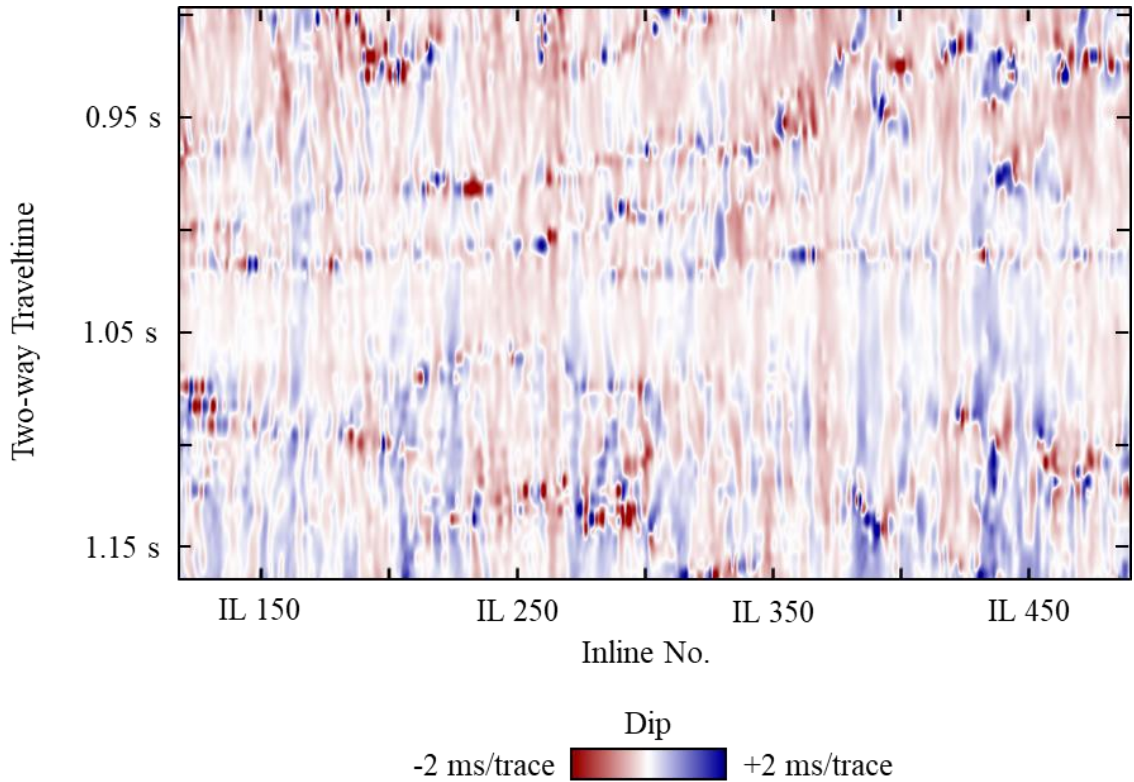

**Fig. 6. The reflector dip in the vertical section of crossline #625 by estimating the orientation of the signed minimum waveform curvature.**

For validating the accuracy of the estimated reflector dip in a more quantitative way, the dip-guided tracking is performed on six reflectors in the vertical section, including two peaks, two troughs, and two zero-crossings (Fig. 7). For each reflector, its picking starts from a pre-defined seed (denoted by the black dots), and uses only the dip to predict the reflector locations at its adjacent traces, without any constraints or adjustments from the waveform similarity. Therefore, the errors in the reflector dip estimates would accumulate from the seed towards both its sides. The less accurate the dip is estimated, the larger misfit we could observe between the pickings and the actual reflectors in the results. As shown in Fig. 7, all the six reflectors are well picked with only subtle misfits (denoted by curves), indicating the accuracy of the dip estimates based on the proposed waveform curvature analysis. Fig. 8 compares the reflector pickings guided by the reflector dips estimated by the proposed approach (in pink) and two traditional approaches, the complex-seismic-trace analysis (in black) and the discrete scanning (in green). Note that both our proposed method and the discrete scanning lead to good pickings of all the three reflectors, but our method has an analytical solution and thus is more computationally efficient. In contrast, although the complex-seismic-trace analysis is comparable in computational efficiency, it undesirably causes dip overestimate that deviates the pickings from the actual locations.

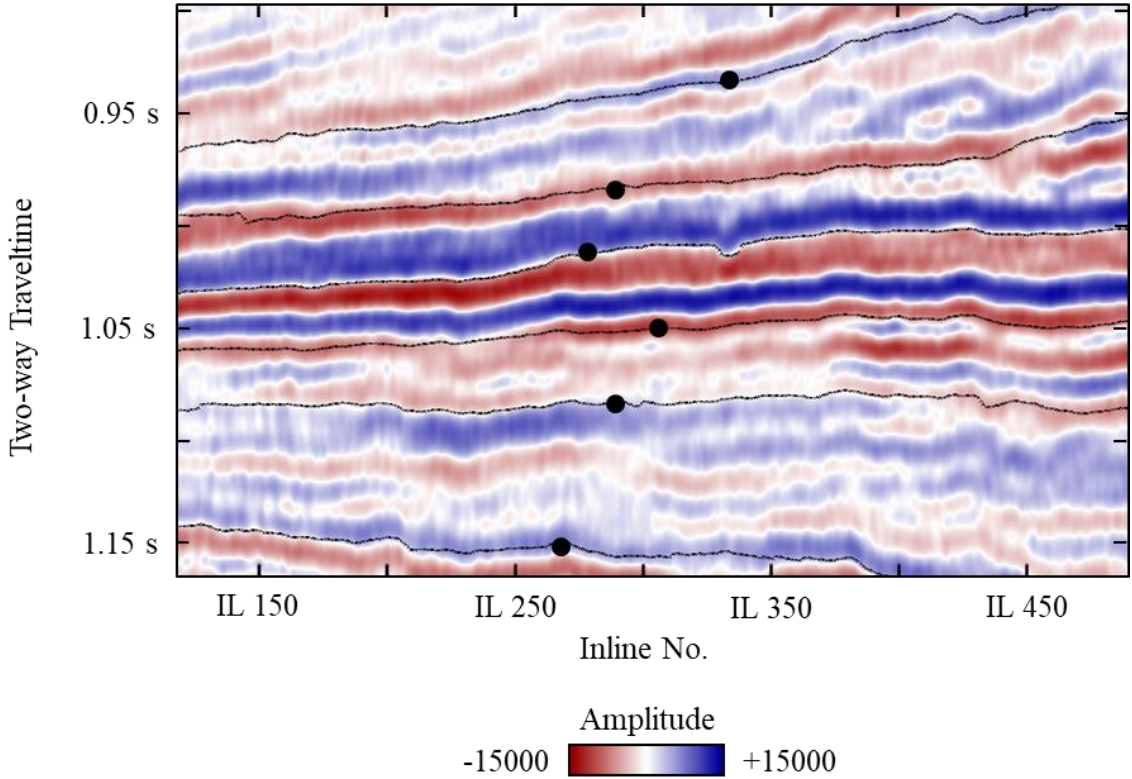

**Fig. 7.** The dip-guided tracking of six reflectors (two peaks, two troughs and two zero-crossings) for validating the accuracy of the reflector dip estimated by the proposed waveform curvature analysis. For each reflector, its picking starts from a defined seed (denoted by the black dots) and uses only the dip to predict the reflector locations at its adjacent traces. Note the subtle misfit between the picking (denoted by the curves) and the actual reflector locations.

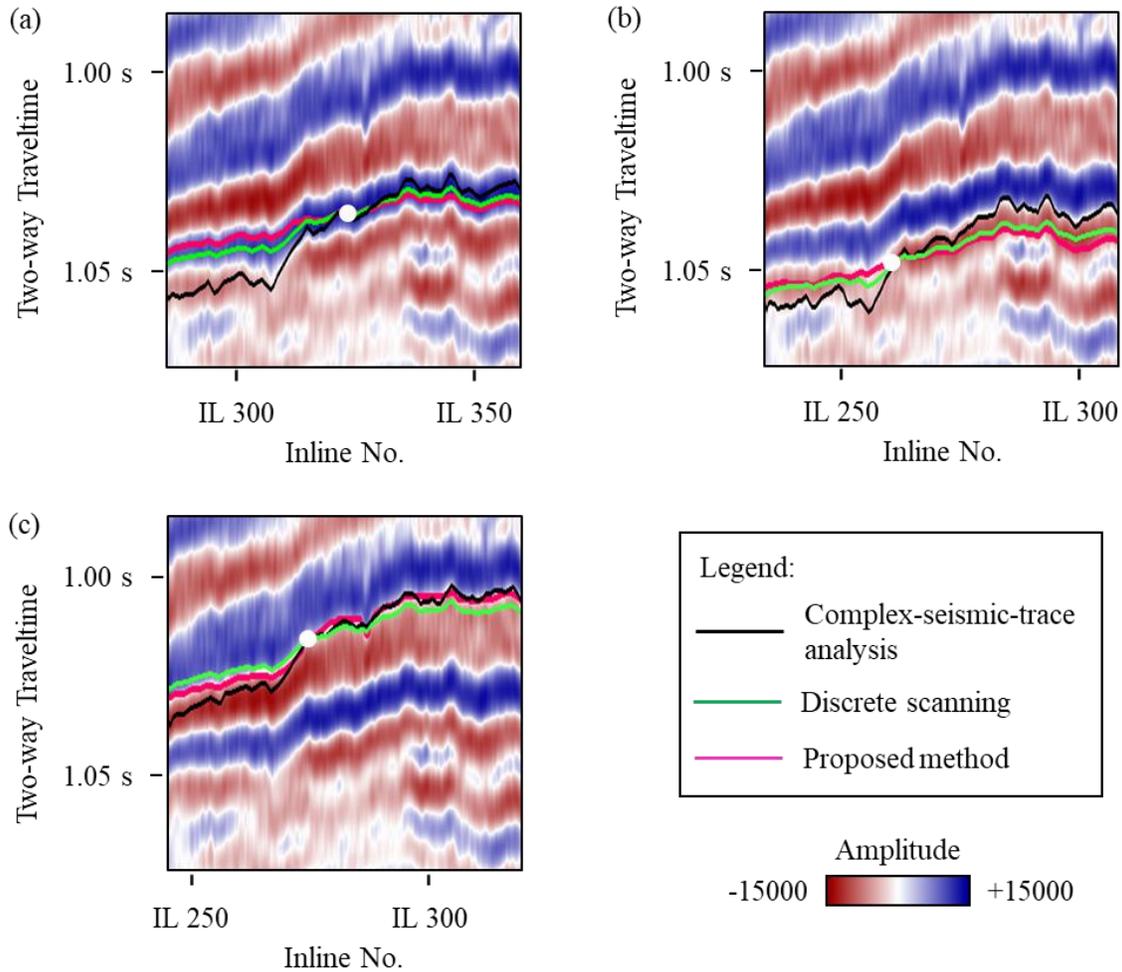

**Fig. 8.** The comparisons of picking three reflections, peak (a), trough (b), and zero-crossing (c) guided by the reflector dips estimated by the proposed method (in pink), the complex-seismic-trace analysis (in black), and the discrete scanning (in green). Note that the picking by the proposed waveform curvature is close to the one by the discrete scanning, indicating its accuracy, whereas the complex-seismic-trace analysis leads to dip overestimates, causing the picking to deviate from the reflections observed in the seismic images.

**Apparent resolution enhancement**

In mathematics, the curvature analysis is capable of capturing the subtle variations in a signal by computing the second-order derivatives, and Fig. 2 has demonstrated such enhancement in the visibility of the weak feature in a seismic trace by computing the 2D waveform curvature. When turning to a seismic section and/or volume, there are numbers of traces that often depict complex subsurface structures with a good lateral continuity in geology. Reliable seismic interpretation is expected to take into account the structural information for avoiding artifacts in attribute analyses and facies mapping (e.g., Marfurt *et al.* 1999; Karimi and Fomel 2015; Marfurt and Alves 2015). The proposed curvature analysis in 3D space offers such a solution for tracking the lateral continuity in structure and thereby retrieving the seismic signals in a way that is more geologically reliable and accurate. As illustrated in Fig. 3, the signed maximum curvature evaluates the seismic signals perpendicularly to the local reflector, instead of simply vertically along the depth/time direction. Therefore, it follows the true direction of geologic deposition and is capable of best preserving the geologic information without requiring any prior knowledge of formation orientations. Fig. 9 displays the signed maximum waveform curvature in the same vertical section of crossline #625. We notice that, compared to the original amplitude (Fig. 4), not only the major reflectors are well delineated without any distortions in their spatial locations, but also the overall apparent resolution is enhanced on better recognizing and interpreting more reflectors that are not discernable from the original amplitude (denoted by circles). Fig. 10 compares the spectrum of the original seismic amplitude and the signed maximum waveform curvature, which clearly demonstrates the improved dominant frequency from 30 Hz to 60 Hz and the enhanced energy in high frequency (>70 Hz). Although such frequency enhancement may not directly improve the true vertical resolution (Chavez-Perez and Centeno-Miranda 2013), it leads to better visibility of weak seismic reflectors and could be particularly helpful for interpreting unconventional, fractured shale reservoirs, in which the reflection is often characterized by low amplitude, low signal-to-noise ratio, and low degree of lateral continuity (Di, Gao and AlRegib 2018).

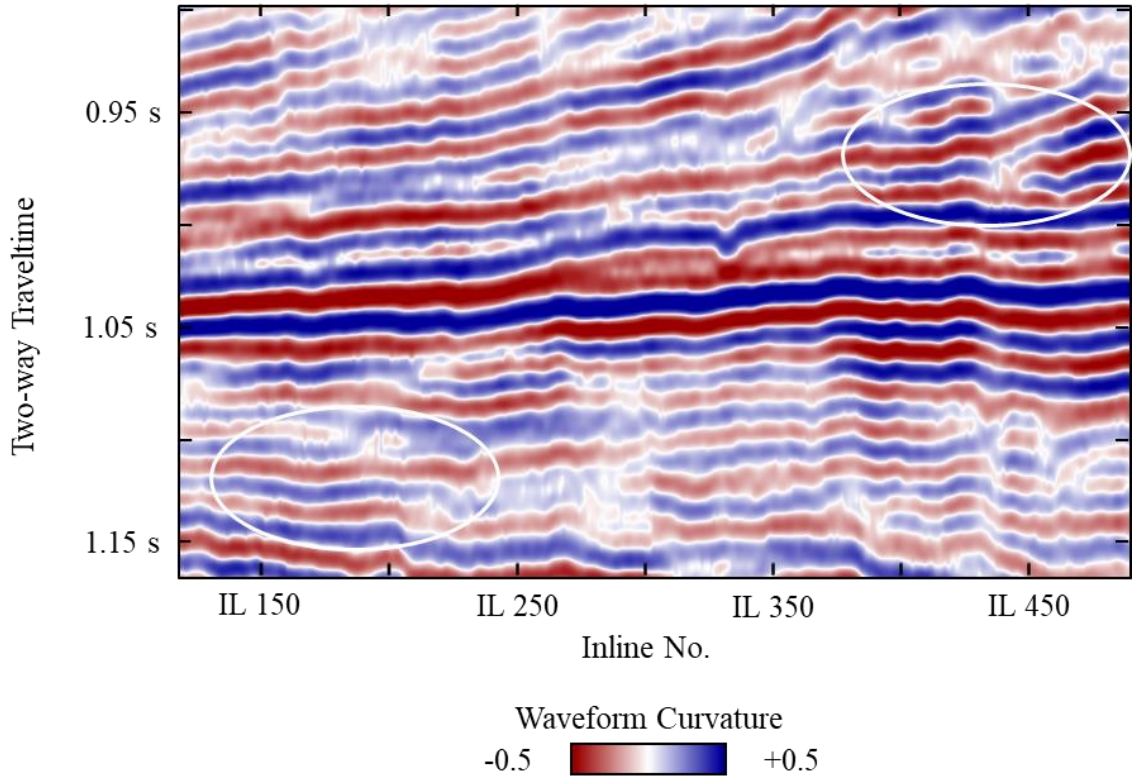

Fig. 9. The signed maximum waveform curvature in the vertical section of crossline #625. Note the enhanced apparent resolution on the weak reflectors in the section (denoted by the circles) compared to the original seismic amplitude (Fig. 4).

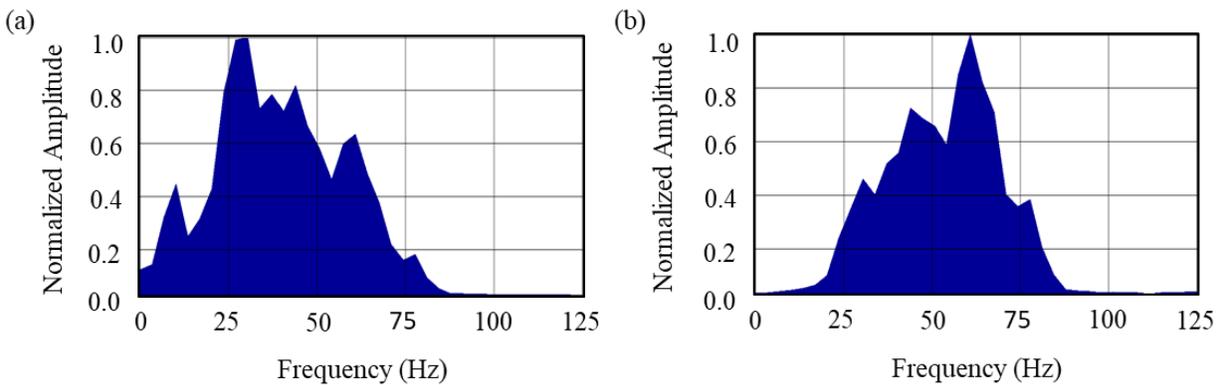

Fig. 10. The comparison of the spectrum of the original seismic amplitude (a) and the signed maximum waveform curvature (b) in the vertical section of crossline #625. Note the increasing of the dominant frequency from 30 Hz to 60 Hz and the enhanced energy in the band of over 70 Hz, both of which are corresponding to the enhanced apparent resolution as observed in Fig. 9.

# APPLICATION EXAMPLE: HORIZON MODELLING

Next, we demonstrate the geophysical applications of the proposed waveform curvature analysis through an example of horizon modelling, which is an important tool for sequence interpretation from a seismic cube and has been the recent focus of geophysical research (de Groot *et al.* 2010; Wu and Hale 2015). A general workflow consists of four steps: first to identify the reflector extrema that represent the deposition events at different geologic time; second to select the seeds for the identified reflector extrema and determine the tracking order among them; then to track each of the horizons with geologic constraints between them; and finally to build a geologic model that obeys all the picked horizons.

**Step 1: Reflector extrema identification**

Horizon interpretations are often placed on a certain type of reflector extrema in the seismic data (e.g., peaks or troughs). In some cases, however, it is not convenient to interpret the target extrema directly from the original seismic images (Fig. 4) containing all reflector types. The reflector decomposition described in the previous section makes it possible to separate each type of reflectors, so that an interpreter could focus on the target type with minimum interference from the rest of the events. These decomposed reflectors can be further thinned by simply extracting the local maximum. Fig. 11 displays the peak extrema and trough extrema in the vertical section of crossline #625 by thinning the decomposed reflectors shown in Fig. 5.

**Step 2: Seed selection and sorting**

Automatic seed selection and sorting is the prerequisite and the key to reliable extraction of multiple reflectors and a horizon volume, and it is often challenging for a computer program to determine the picking priorities between these reflectors, especially in the areas of geologic complexities (Yu, Kelley and Mardanova 2011). Based on the reflector extrema identified in Fig.

11, the step of seed selection and sorting can then be completed in two steps (Fig. 12). First, at a seismic trace, the computer retrieves only the samples whose waveform curvature is larger than a defined threshold, and places one seed on each sample. Second, these seeds are sorted by their curvature magnitude from high to low (denoted by dot size). The larger the magnitude is, the higher priority the seed has in the step of horizon picking. The sorting scheme is based on an assumption that the significance of a reflection event is proportional to its reflection intensity as well as the associated magnitude of waveform curvature, which is generally consistent with the sequence of manual horizon interpretation in a seismic volume.

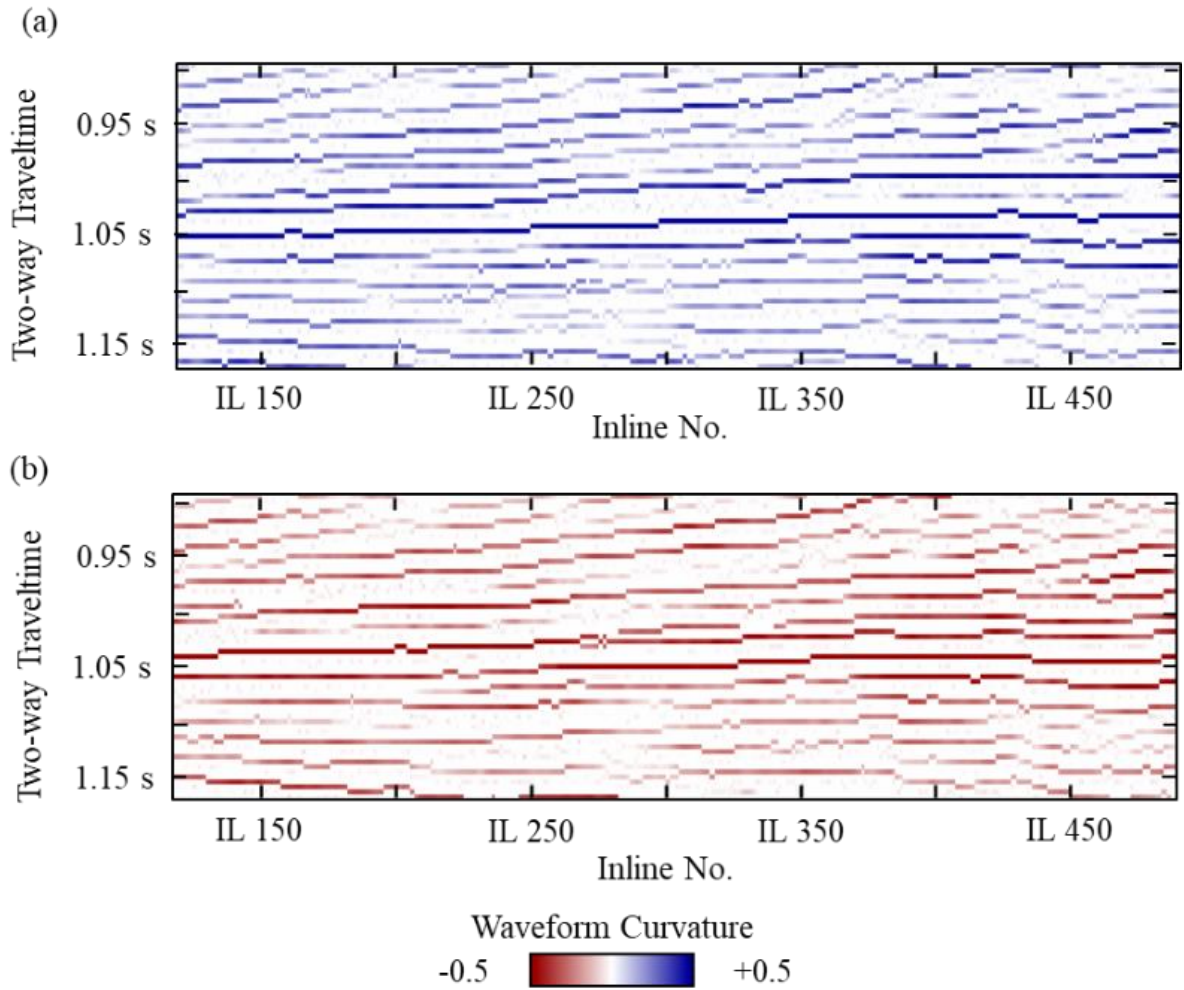

**Fig. 11. The peak extrema (a) and trough (b) extrema in the vertical section of crossline #625 by thinning the decomposed reflectors in Fig. 5.**

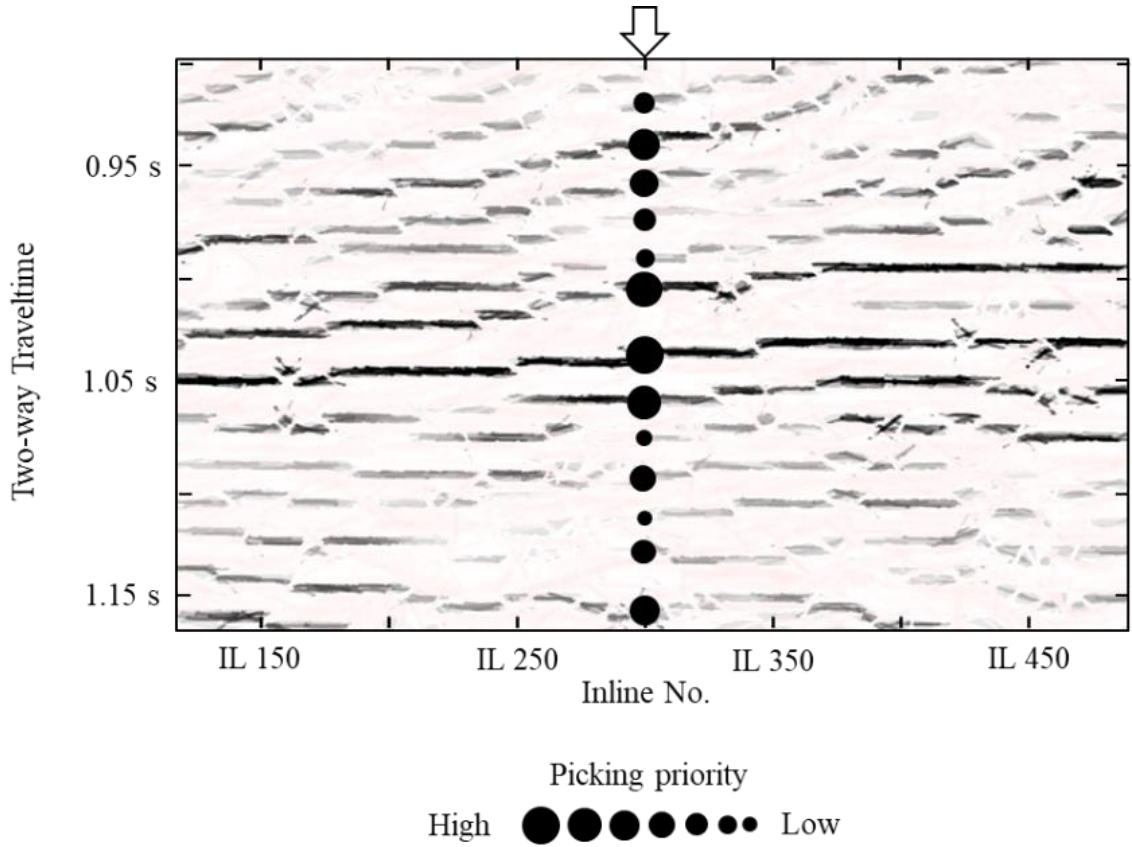

**Fig. 12.** The automatic seed selection and sorting for picking the peak reflectors in the vertical section of crossline #625. At a seismic trace (denoted by the arrow), a seed is placed on each of the peak extrema, and these seeds are then sorted based on the magnitude of the waveform curvature to determine the priority of picking the peaks (denoted by dot size).

**Step 3: Horizon tracking**

Performing horizon tracking on the sorted seeds leads to a volume of all horizons. For automatic horizon tracking, various horizon trackers have been developed and implemented (e.g., Harrigan *et al.* 1992; Leggett, Sandham and Durrani 1996; Huang 1997; Faraklioti and Petrou 2004; Patel *et al.* 2010; Yu *et al.* 2011; Di *et al.* 2018). This study uses the dip vector-guided method (Di *et al.* 2018) to grow the tracking from the given seed through the entire survey, and the dip vector is prepared using the proposed waveform curvature analysis (Fig. 6). Fig. 13 displays the 3D view of the ten peak reflectors extracted from the F3 dataset. For quality control, all trackings are then clipped to the vertical section of crossline #625 (Fig. 14). Good matches are noticed between the tracked ten horizons and the original seismic images, which helps verify the accuracy of horizon tracking and ensure the reliability of horizon modelling in step 4.

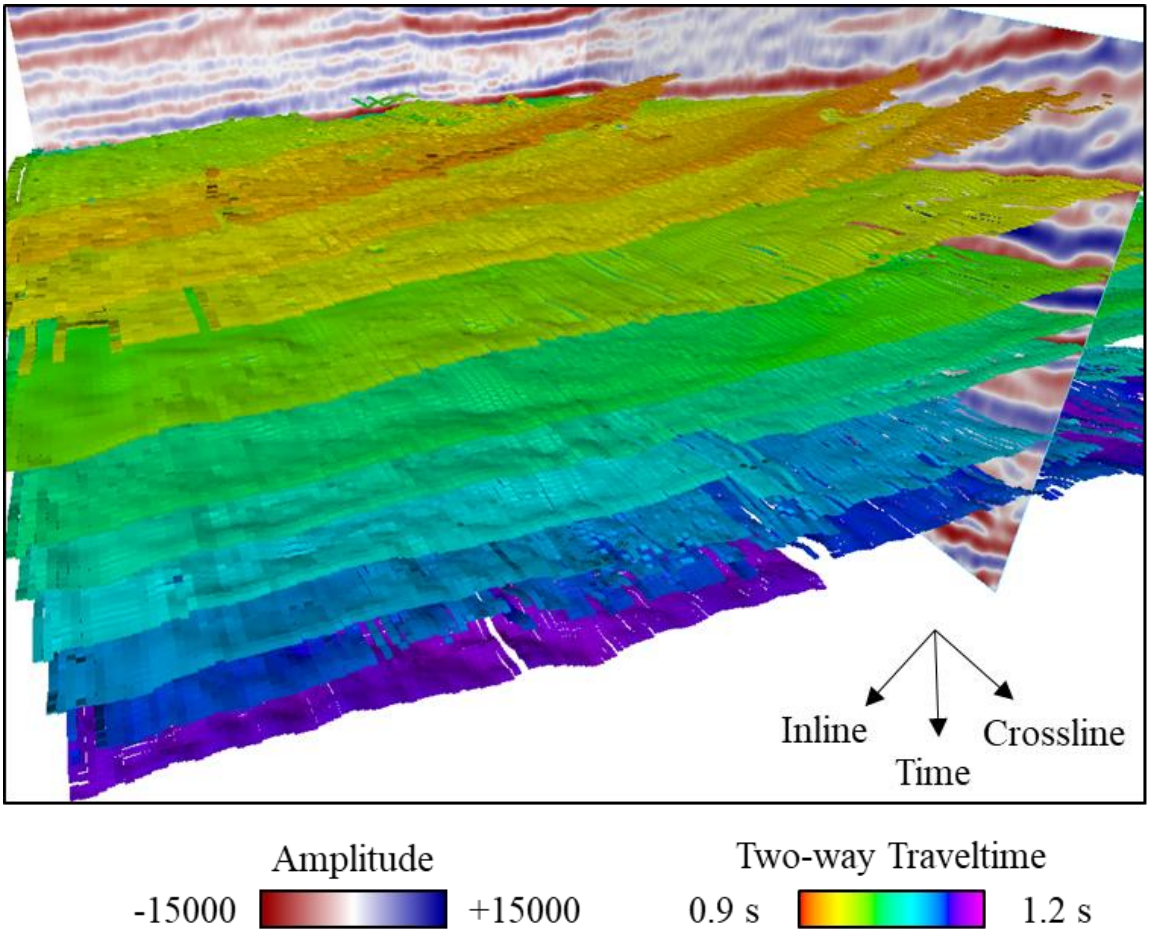

**Fig. 13. The 3D view of the ten peak reflectors automatically extracted from the F3 seismic dataset.**

**Step 4: Horizon modelling**

With the prepared ten peak reflectors as input, horizon modelling is then performed to provide us with a horizon model that simulates the deposition process in geology in the F3 area. Fig. 15 displays the 2D view of such a model by clipping it to the vertical section of crossline #625. The colors are randomly assigned for the sole purpose of differentiating these reflectors.

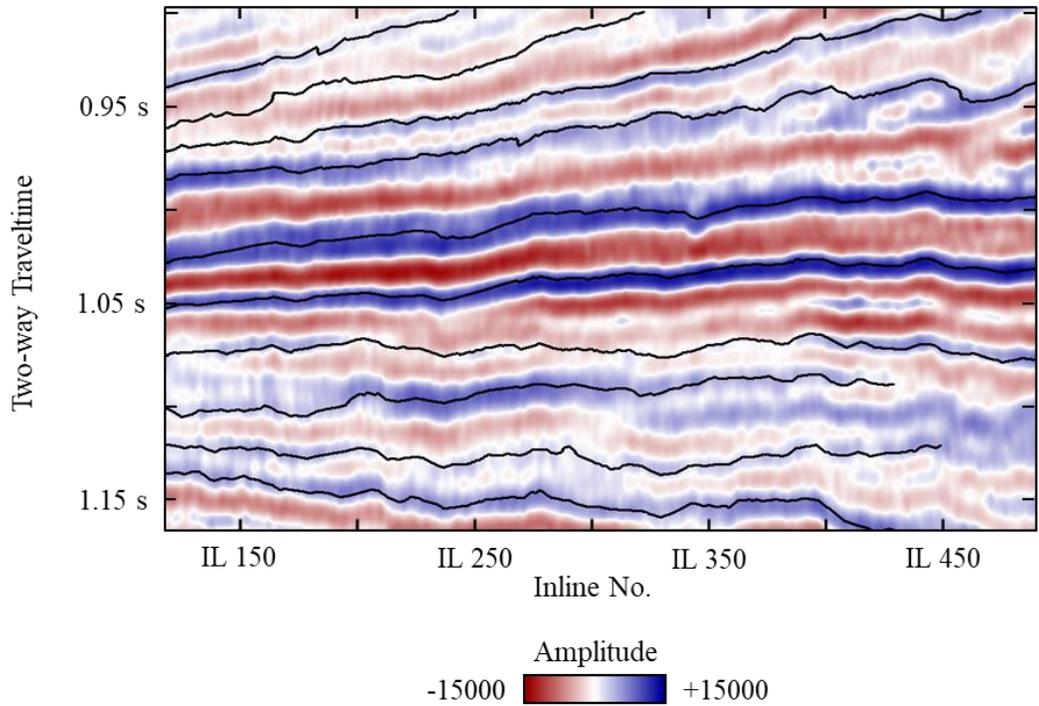

**Fig. 14.** The clipping of the ten peak reflectors to the vertical section of crossline #625. Note the good match between the pickings and the original seismic image.

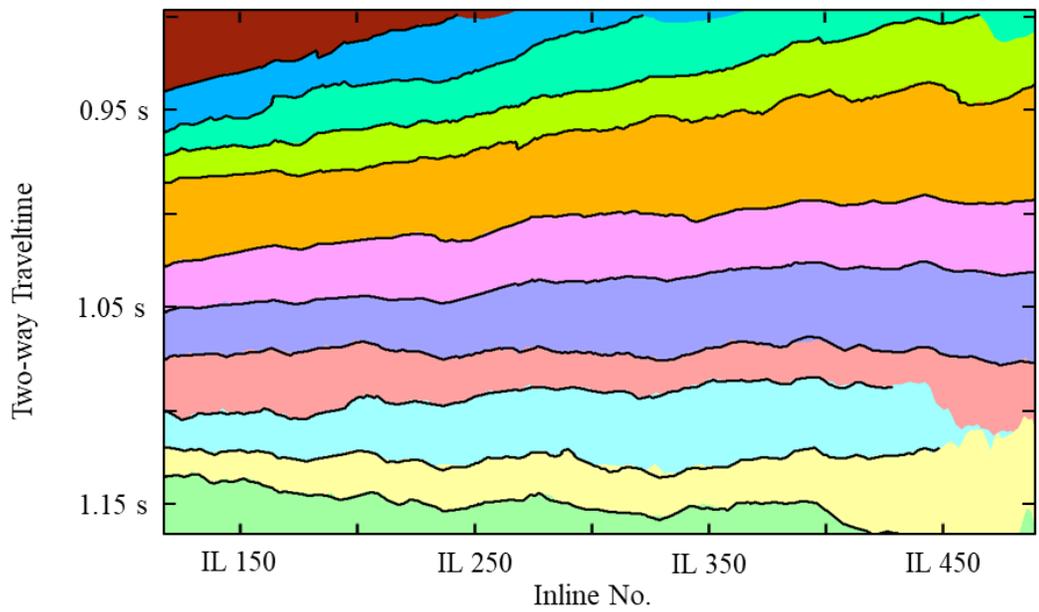

**Fig. 15.** The 2D view of the horizon model built using the picked ten peak reflectors (Fig. 13). The colors are randomly assigned for the sole purpose of differentiating these reflectors.

## DISCUSSIONS

Although the proposed waveform curvature analysis has several helpful interpretational implications as discussed above, interpreters should bear in mind that the seismic signals are reformed in a certain way by the curvature operator. Correspondingly, the original reflection patterns have been changed, and misinterpretation would arise without taking into account the impacts of the curvature analysis. Therefore, it is necessary to investigate the possible limitations of the proposed waveform curvature, and we list the major ones in the following.

**Sensitivity to local high-frequency components**

When applied to a seismic dataset, the waveform curvature evaluates how the seismic waveform varies from one sample to its neighbors. Thereby, the magnitude of the estimated waveform curvature is dependent on not only the reflection amplitude but also the waveform frequency. Qualitatively, for a given waveform frequency, the stronger the reflection is, the more apparently the amplitude varies, and correspondingly, the waveform curvature is estimated larger. Similarly, for a given waveform amplitude, the higher the frequency is, the faster the amplitude varies, and correspondingly, the waveform curvature is also estimated larger. For quantifying the impacts from both elements, we take the common Ricker wavelet as example, which is often defined as

$$w(t) = A \cdot \left(1 - 2\pi^2 f_M^2 t^2\right) e^{-\pi^2 f_M^2 t^2} , \tag{3}$$

in which $A$ represents the amplitude and $f_M$ represents the dominant frequency. Substituting eq. 3 into eq. 1 provides us with the associated waveform curvature $k(t, w)$. For simplification, we consider the wavelet peak at zero time ($t = 0$), and eq. 4 represents the relationship between its curvature $k(0, w)$ and the wavelet amplitude/frequency:

$$k(0, w) = -6 \cdot A \cdot \pi^2 f_M^2 . \tag{4}$$

It informs us that the waveform curvature is linearly related to the amplitude $A$, but quadratically related to the dominant frequency $f_M$, implying that the waveform curvature analysis is more sensitive to the high-frequency components in seismic waveforms. Therefore, the local components with low amplitude but high frequency in a seismic signal would be exaggerated after computing its waveform curvature. Such exaggeration may interfere with other reflectors, change local reflection patterns, and add difficulties for reflector interpretation.

Fig. 16 shows such interference observed in the F3 dataset after applying the waveform curvature analysis. The trough reflector (denoted by the arrow) is visually connected to another trough in the image of waveform curvature (denoted by the circle) (Fig. 16b), due to its relatively high frequency related to the reflection discontinuity. However, it has low amplitude, and such pattern is not observed from the original amplitude (Fig. 16a).

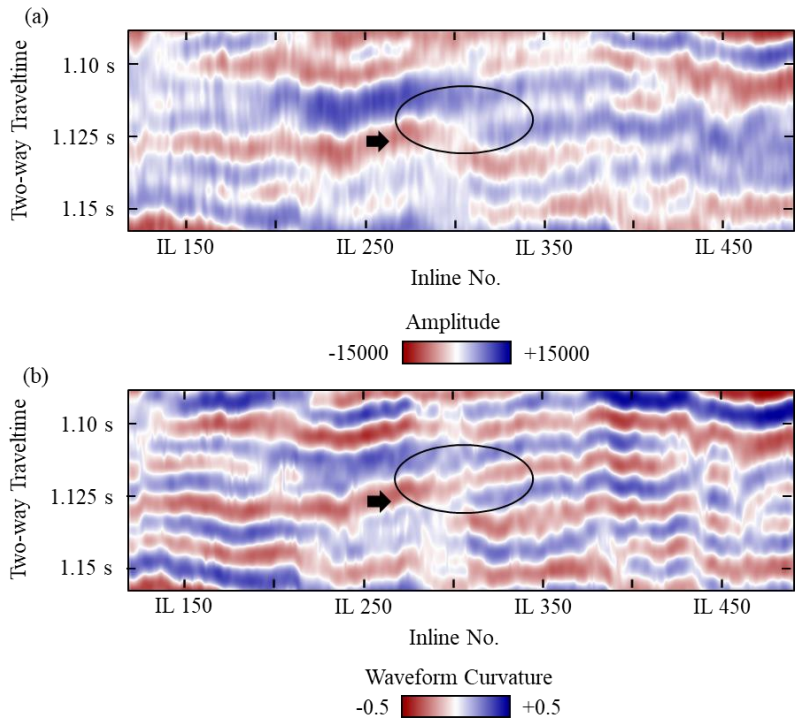

**Fig. 16. The sensitivity of the proposed waveform curvature analysis to local high-frequency components. The trough reflector is visually connected to another trough (denoted by the circle) after applying the waveform curvature analysis (b), due to its relatively high frequency related to the local reflection discontinuity. However, it has low amplitude, and such pattern is not observed from the original amplitude (a).**

**Sensitivity to time window size**

The computation of seismic waveform curvature requires a time analysis window for quantifying the derivative terms used in eqs. 1 and 2, and thus the accuracy of the waveform retrieval determines the accuracy of curvature analysis. Take the first derivatives for example. As shown in Fig. 17, in a small window (denoted by the red arrow) the waveform amplitude distribution is nearly linear, and the corresponding waveform curvature could be accurately estimated. However, with the window size increasing (denoted by the yellow/green/blue arrows), the non-linear nature of waveforms is incorporated into the analysis window, leading to inaccurate waveform retrieval as well as underestimated waveform curvature. In an extreme case, when the window is larger than half the waveform period, the amplitude variation would be close to zero (denoted by the blue arrow). Similarly, such sensitivity is even higher for the second derivatives.

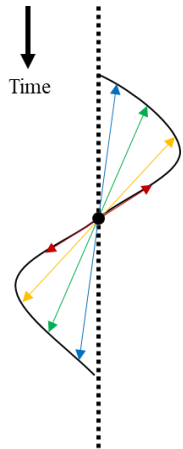

**Fig. 17. The diagram for illustrating the sensitivity of waveform curvature analysis to time window size. Due to the non-linear nature of a waveform, a small window (denoted by the red arrow) best represents the waveform variation. With the window increasing (denoted by the yellow/green/blue arrows), such non-linear distribution leads to inaccurate curvature estimation, especially when the window size is larger than half the waveform period (denoted by the blue arrow).**

Fig. 18 compares the waveform curvatures estimated from the same section with various time windows. The curvature underestimate becomes more and more apparent with the window size increasing from 4 *ms* (Fig. 18a) to 16 *ms* (Fig. 18d). Therefore, it is recommended that the window size is less than half the waveform period to avoid underestimation. In most cases, the waveform period is unknown, and thus a size of double the sampling rate can tentatively be used and then gradually increased until the result fits the interpretational purposes.

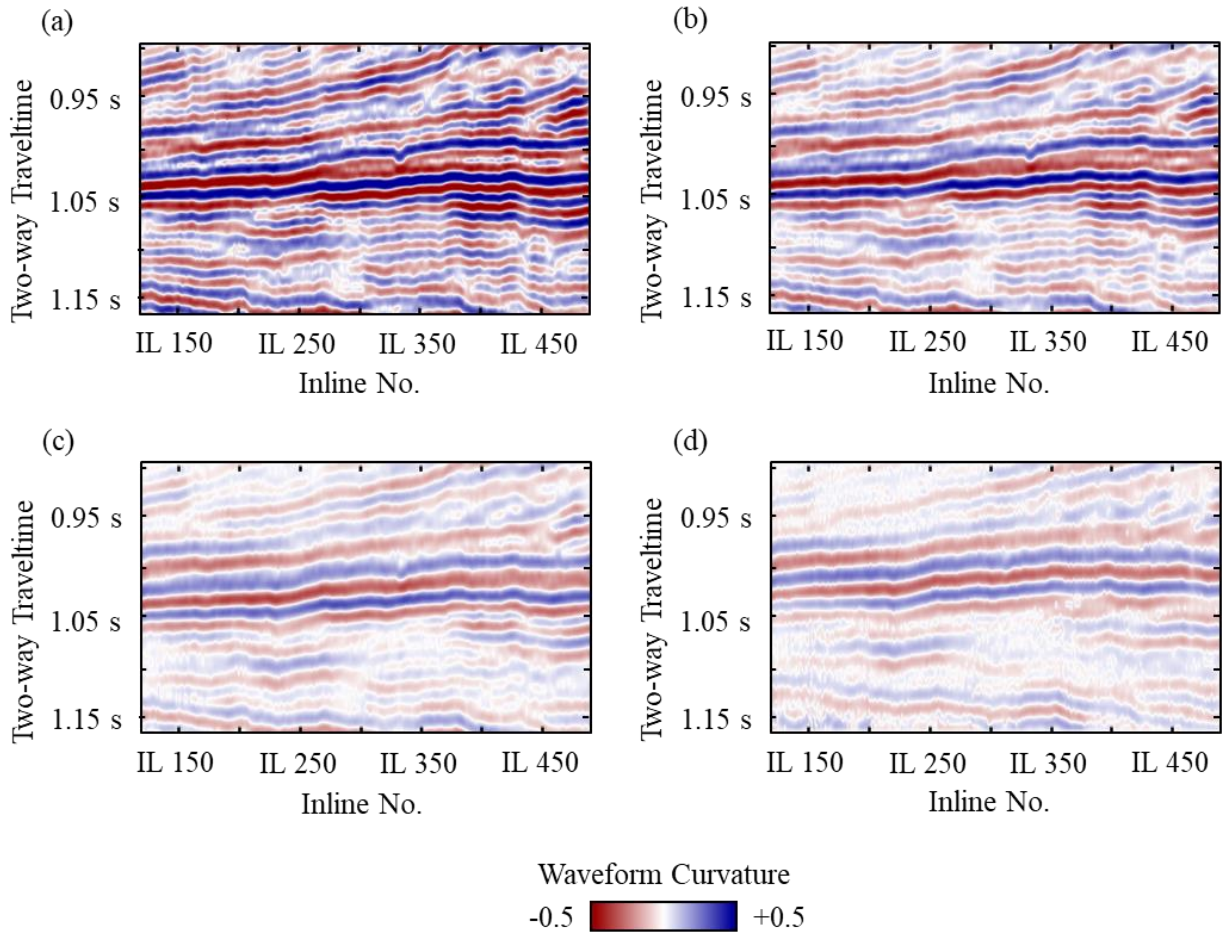

Fig. 18. The sensitivity of the proposed waveform curvature analysis to time window size of 4 *ms* (a), 8 *ms* (b), 12 *ms* (c) and 16 *ms* (d) samples. Note that with the analysis window increasing, the non-linear distribution of waveform amplitude (Fig. 17) causes underestimates of the waveform curvature, particularly in (c) and (d).

**Sensitivity to noises**

Similar to the traditional geometric curvature, as the second-derivative-related operation, the waveform curvature analysis also runs the risk of exaggerating random noises that are present in seismic waveforms. Such exaggeration may appear as artifacts in the images of waveform curvature, interrupt the reflector continuity, and increase the difficulties of seismic interpretation (Fig. 19a). Therefore, structure-preserving smoothing is strongly recommended to suppress local noises before performing the proposed waveform curvature analysis. Fig. 19b displays the corresponding result from the dip-guided median filter with window size of 5 inlines by 5 crosslines by 8 *ms*. Meanwhile, while reading the generated waveform curvature maps, the interpreters are recommended taking into account their experience and additional geologic information (e.g., structure orientation) for differentiating the revealed subtle structures from the introduced artifacts. Future work is expected to quantify such sensitivity and developing new approaches for automated artifact detection.

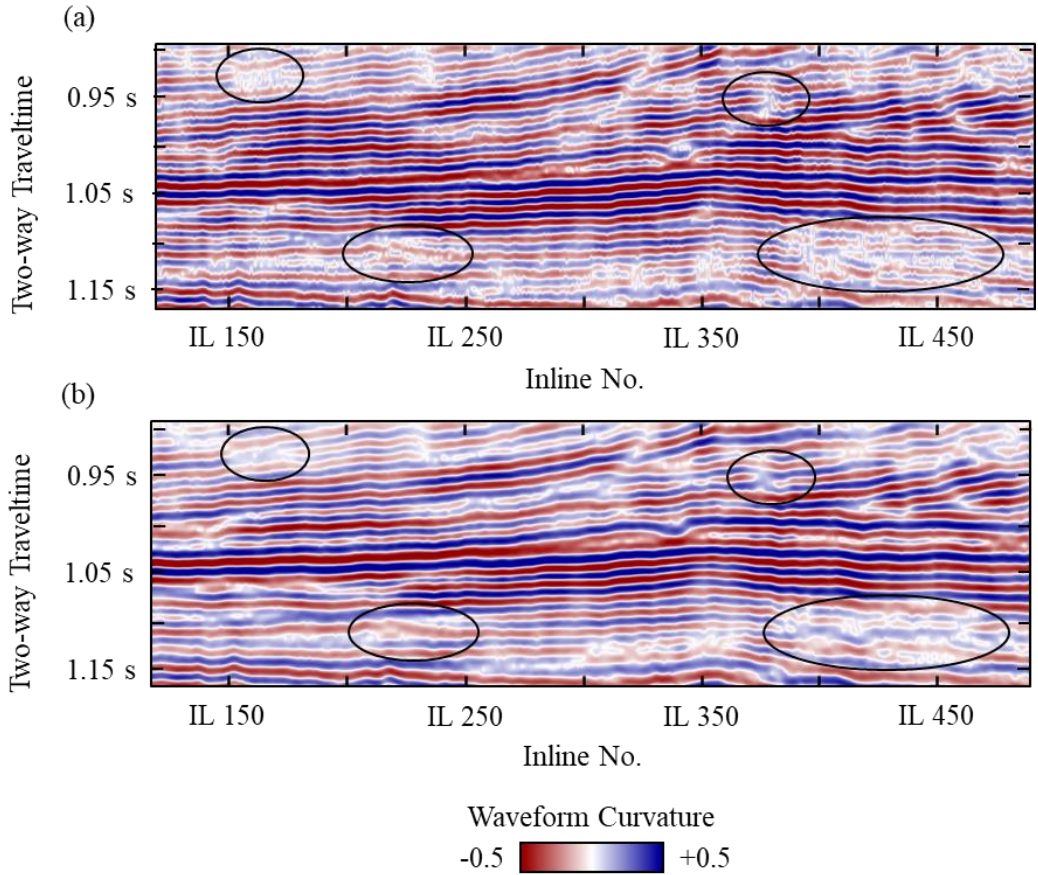

**Fig. 19. The sensitivity of the proposed waveform curvature analysis to seismic noises. In the presence of noises, the curvature operator exaggerates and incorporate them into the results, which not only breaks the reflector continuity but also increase the interpretation difficulties (denoted by the circles in a). Structure-oriented smoothing, such as the dip-guided median filter, is recommended to help reduce such sensitivity (b).**

# CONCLUSIONS

This study has extended the traditional curvature analysis to the seismic waveforms and investigated the associated implications for assisting seismic interpretation. In general, for a given seismic section, the waveform curvature retrieves the signals perpendicularly to the reflectors, which follows the true direction of geologic deposition and thereby is most accurate and reliable in geology. Among all possible curvatures from a seismic section and/or volume, the signed maximum curvature provides an enhanced apparent resolution, from which the weak reflectors are better delineated and become more recognizable for interpretation. The signed minimum curvature offers an analytical approach for finding the orientation of least waveform variation (or maximum reflector continuity), which allows estimating the reflector dip in a more accurate and efficient manner. Integrated with the complex seismic trace analysis, the maximum and minimum curvatures make it possible to decompose a seismic profile by the four major types of reflection events (peaks, troughs, s-crossings, and z-crossings). Applications to the F3 seismic dataset over the Netherlands North Sea verifies the impacts and applications of the proposed waveform curvature analysis on automated volumetric horizon interpretation and modelling from 3D seismic data.

# ACKNOWLEDGEMENTS

This work is supported by the Center for Energy and Geo Processing at Georgia Tech and King Fahd University of Petroleum and Minerals. Thanks go to the Editor Dr. Noalwenn DUBOS-SALLEE, the Associate Editor Dr. Herve Chauris, and the two anonymous reviewers for their helpful insights for improving the quality and clarity of the manuscript. We thank the OpendTect

Open Seismic Repository (opendtect.org/osr) for providing the F3 dataset over the Netherlands North Sea and Schlumberger for providing an academic Petrel license.